\documentclass[twocolumn,showpacs,pre,amsmath,amssymb]{revtex4}

\usepackage{graphicx}
\usepackage{dcolumn} 
\usepackage{bm}      
\usepackage{latexsym}
\usepackage{color}

\begin{document}

\title{Low magnetic Prandtl number dynamos with helical forcing}

\author{Pablo D. Mininni$^1$ and David C. Montgomery$^2$}

\affiliation{$^1$ National Center for Atmospheric 
Research, P.O. Box 3000, Boulder, Colorado 80307}
\affiliation{$^2$ Dept. of Physics and Astronomy,
Dartmouth College, Hanover, NH 03755} 

\date{\today}

\begin{abstract}
We present direct numerical simulations of dynamo action in a 
forced Roberts flow. The behavior of the dynamo is followed as 
the mechanical Reynolds number is increased, starting from the 
laminar case until a turbulent regime is reached. The critical
magnetic Reynolds for dynamo action is found, and in the turbulent 
flow it is observed to be nearly independent on the magnetic Prandtl 
number in the range from $\sim 0.3$ to $\sim 0.1$. Also the 
dependence of this threshold with the amount of mechanical helicity 
in the flow is studied. For the different regimes found, the configuration 
of the magnetic and velocity fields in the saturated steady state are 
discussed.
\end{abstract}

\maketitle

\section{\label{sec:intro}INTRODUCTION}

In a previous publication \cite{Ponty05}, a driven turbulent 
magnetohydrodynamic (MHD) dynamo was studied numerically, within the 
framework of rectangular periodic boundary conditions. The emphasis 
was on the dynamo's behavior as the magnetic Prandtl number $P_M$ 
(ratio of kinematic viscosity to magnetic diffusivity) was lowered. 
As $P_M$ is lowered at fixed viscosity, the magnetofluid becomes more 
resistive than it is viscous, and it is intuitively apparent that 
magnetic fields will be harder to excite by mechanical motions. The 
principal result displayed in Ref. \cite{Ponty05} was a curve of critical 
magnetic Reynolds number, $R_M^c$, as a function of $P_M^{-1}$, at 
fixed kinetic energy. The (turbulent) kinetic energy was the result 
of an external mechanical forcing of the Taylor-Green type (hereafter, 
``TG''), a geometry well known to be efficient at the rapid generation 
of small scales in the fluid flow \cite{Taylor37}. Ref. \cite{Ponty05} 
contains a lengthy bibliography of its antecedents, not all of which 
will be listed again here.

The TG geometry injects no net mechanical helicity into the flow. In 
the long history of the dynamo problem, mechanical helicity has been 
seen often to be an important ingredient for dynamo action, and it is 
the intent of this present paper to consider a helically-forced dynamo 
in the same spirit as in Ref. \cite{Ponty05}, to see what changes occur 
relative to the TG flow, further properties of which were displayed 
in a subsequent astrophysical paper \cite{Mininni05a}.

A natural candidate for a highly helical velocity field is what has 
come to be called the ``Roberts flow'' \cite{Roberts72,Dudley89}. 
This flow shares some similarities with the dynamo experiments 
of Riga and Karlsruhe \cite{Gailitis01,Stieglitz01}. In a 
pioneering paper \cite{Feudel03}, Feudel et al. characterized 
mathematically various magnetic-field-generating instabilities that 
a forced Roberts flow can experience. The present paper expands these 
investigations, while discussing numerical simulation results for 
magnetic excitations in the mechanically turbulent regime, with an 
emphasis on the nonlinearly-saturated magnetic field configuration. 
As in Ref. \cite{Feudel03}, we will force the system at nearly the 
largest scale available in the periodic domain. As a result, magnetic 
fields will be only amplified at scales smaller than the energy containing 
scale of the flow. The behavior of the large-scale dynamo (i.e. when 
magnetic perturbations are amplified at scales larger than the energy 
containing eddies) as $P_M$ is varied will be studied in a future work.

Section \ref{sec:equations} lays out the dynamical equations and 
definitions and describes the methodology to be used in the numerical 
study. Section \ref{sec:results} presents results and compares some 
of them with the corresponding TG results. Section \ref{sec:summary} 
summarizes and discusses what has been presented, and points in 
directions that we believe the results suggest.

\section{\label{sec:equations}MATHEMATICAL FRAMEWORK AND METHODOLOGY}

In a familiar set of dimensionless (``Alfv\'enic'') units the 
equations of magnetohydrodynamics to be solved are:
\begin{eqnarray}
\frac{\partial {\bf v}}{\partial t} + {\bf v \cdot \nabla v} &=& 
    -{\bf \nabla} {\mathcal P} + {\bf j \times B} + \nu \nabla^2 {\bf v} + 
    {\bf f} , \label{eq:momentum} \\
\frac{\partial {\bf B}}{\partial t} + {\bf v \cdot \nabla B} &=&
    {\bf B \cdot \nabla v} + \eta \nabla^2 {\bf B} ,
    \label{eq:induc}
\end{eqnarray}
with ${\bf \nabla \cdot v} = 0$, ${\bf \nabla \cdot B} = 0$. ${\bf v}$ 
is the velocity field, regarded as incompressible (low Mach number). 
${\bf B}$ is the magnetic field, related to the electric current density 
${\bf j}$ by ${\bf \nabla \times B} = {\bf j}$. ${\mathcal P}$ is the 
normalized pressure-to-density ratio, obtained by solving the Poisson 
equation for it that results from taking the divergence of Eq. 
(\ref{eq:momentum}) and using the incompressibility condition 
${\bf \nabla \cdot v} = 0$. In these units, the viscosity $\nu$ and 
magnetic diffusivity $\eta$ can be regarded as the reciprocals of 
mechanical Reynolds numbers and magnetic Reynolds numbers respectively, 
where these dimensionless numbers in laboratory units are 
$R_V = LU/\nu_{lab}$, $R_V = LU/\eta_{lab}$. Here $U$ is a typical 
turbulent flow speed (the r.m.s. velocity in the following sections), 
$L$ is a length scale associated with its spatial variation (the 
integral length scale of the flow), and $\nu_{lab}$, $\eta_{lab}$ are 
kinematic viscosity and magnetic diffusivity, respectively, 
expressed in dimensional units. The external forcing function ${\bf f}$ 
is to be chosen to supply kinetic energy and kinetic helicity and to 
maintain the velocity field ${\bf v}$.

For ${\bf f}$, we choose in this case the Roberts flow 
\cite{Roberts72,Feudel03}:
\begin{equation}
{\bf f} = - \nu \nabla^2 {\bf v}_R = 2 \nu {\bf v}_R
\label{eq:laminar}
\end{equation}
where
\begin{equation}
{\bf v}_R = (g \sin x \cos y , -g \cos x \sin y, 2 f \sin x \sin y ) .
\label{eq:Roberts}
\end{equation}
The coefficients $f$ and $g$ are arbitrary and their ratio determines 
the extent to which the flow excited will be helical. The ratio 
$f = g/\sqrt{2}$ is maximally helical for a given kinetic energy, 
and the case $f/g \to 0$ is a (two-dimensional) non-helical excitation. 
We have concentrated primarily upon the cases $f = g$ (following Feudel 
et al. \cite{Feudel03}) and $f = g/\sqrt{2}$. No dynamo can be expected 
unless $|f/g| > 0$.

We impose rectangular periodic boundary conditions throughout, using 
a three-dimensional periodic box of edge $2\pi$, so that the fundamental 
wavenumber has magnitude 1. All fields are expanded as Fourier series, 
such as
\begin{equation}
{\bf v} = {\bf v}({\bf x},t) = \sum_{\bf k}{{\bf v}({\bf k},t) 
    \exp (i {\bf k \cdot x})} 
\end{equation}
with ${\bf k \cdot v}({\bf k},t) = 0$. The Fourier series are truncated at 
a maximum wavenumber $k_{max}$ that is adequate to resolve the smallest 
scales in the spectra. The method used is the by-now familiar 
Orzag-Patterson pseudospectral method \cite{Orszag72a,Orszag72b,Canuto}. 
The details of the parallel implementations of the Fast Fourier 
Transform can be found in Ref. \cite{Gomez05}.

The forcing function (\ref{eq:Roberts}) injects mechanical energy at a 
wavenumber $|{\bf k}| = \sqrt{2}$, which leaves very little room in the 
spectrum for any back-transfer of helicity ($|{\bf k}|=1$ is the 
only possibility). The phenomena observed will therefore be 
well-separated from those where an ``inverse cascade'' of magnetic 
helicity is expected to be involved. Rather, a question that can be 
answered (in the affirmative, it will turn out) is, To what extent 
does the presence of mechanical helicity in the large scales make it 
easier to excite magnetic fields through turbulent dynamo action?

Equations (\ref{eq:laminar}) and (\ref{eq:Roberts}) define a steady 
state solution of Eqs. (\ref{eq:momentum}) and (\ref{eq:induc}), with 
${\bf B} = 0$. It is to be expected that for large enough $\nu$ and 
$\eta$, this solution will be stable. As the transport coefficients 
are decreased, it will be the case that the flow of Eq. (\ref{eq:Roberts}) 
can become unstable, either purely mechanically as an unstable 
Navier-Stokes flow, or magnetically as a dynamo, or as some combination 
of these. Thus rather complex scenarios can be imagined as either of 
the Reynolds numbers is raised.

In the following, the emphasis will be upon discovering thresholds in 
$R_M$ at which dynamo behavior will set in as $R_V$ is raised, then 
further computing the nonlinear regime and saturation of the magnetic 
excitations once it does. The ``growth rate'' $\sigma$ can be defined 
as $\sigma = d \ln(E_M) / dt$, where 
$E_M = \sum_{\bf k}{|{\bf B}({\bf k},t)|^2 /2}$ is the total magnetic 
energy. The appearance of a positive $\sigma$ for initially very small 
$E_M$ is taken to define the critical magnetic Reynolds number $R_M^c$ 
for the onset of dynamo action. $\sigma$ is typically expressed in units 
of the reciprocal of the large-scale eddy turnover time $L/U$ where 
$U$ is the r.m.s. velocity ($U = \left<u^2\right>^{1/2}$, and the 
brackets denote spatial average), and $L$ is the integral length scale, 
\begin{equation}
L = 2 \pi \sum_{\bf k}{k^{-1} |{\bf u}({\bf k},t)|^2} \bigg/ 
    \sum_{\bf k}{|{\bf u}({\bf k},t)|^2} .
\label{eq:integral}
\end{equation}

In the next Section, we describe the results of the computations for 
both the ``kinematic dynamo'' regime [where ${\bf j \times B}$ is 
negligible in Eq. (\ref{eq:momentum})], and for full MHD where the 
Lorentz force modifies the flow.

\section{\label{sec:results}DYNAMO REGIMES FOR THE ROBERTS FLOW}

\begin{figure}
\includegraphics[width=9cm]{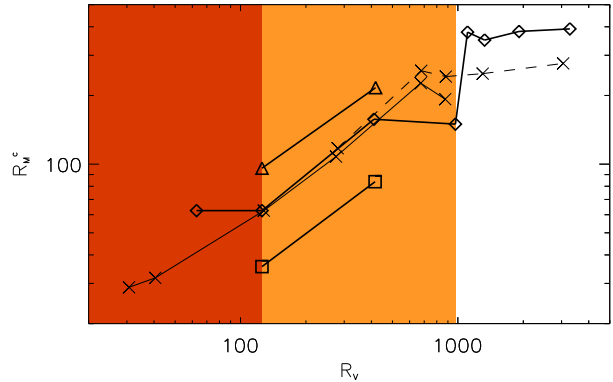}
\caption{(Color online) Critical magnetic Reynolds $R_M^c$ as a 
     function of $R_V$ for different Roberts flows (thick lines): $f=g$ 
     ($\Diamond$), $f=g/\sqrt{2}$ ($\Box$), $f=g/0.77$ ($\triangle$). 
     The dark (red) area corresponds to the region where the Roberts 
     flow is hydrodynamically stable. For a comparison with Ref. 
     \cite{Feudel03}, our Reynolds numbers should be divided by $2\pi$. 
     The light (orange) area corresponds to the region of hydrodynamic 
     oscillations, while the white area corresponds to the turbulent 
     regime. The thin lines connected with crosses are shown as a 
     reference and correspond to the threshold for dynamo instability in 
     Taylor-Green flow \cite{Ponty05}: DNS (solid line) and $\alpha$-model 
     (dashed line).}
\label{fig:crmm}
\end{figure}

We introduce the results for the Roberts flow through a plot of 
the threshold values of critical magnetic Reynolds number $R_M^c$ 
vs. mechanical Reynolds number $R_V$ (Fig. \ref{fig:crmm}). All 
Reynolds numbers have been computed using the integral scale for 
the velocity field [Eq. (\ref{eq:integral})], averaged over the 
duration of the steady state in the hydrodynamic simulation. For 
this same time interval, an overall normalization factor has been 
multiplied by Eq. (\ref{eq:laminar}) to make the r.m.s. velocity 
$U$ turn out to have a time-averaged value of about 1.

\begin{figure}
\includegraphics[width=9cm]{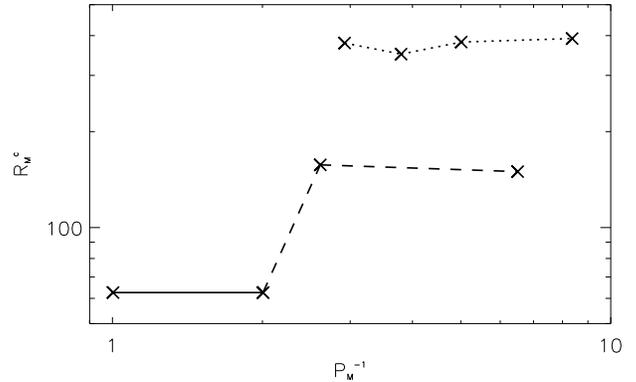}
\caption{Critical magnetic Reynolds $R_M^c$ as a function of $P_M^{-1}$ 
     for the Roberts flow with $f=g$ (thick lines). The solid line 
     corresponds to the laminar regime, the dashed line to the periodic 
     flow, and the dotted line to the turbulent regime. (The 
     double-valuedness results from the effects of two different 
     values of $R_V$.)}
\label{fig:crll}
\end{figure}

Figure \ref{fig:crmm} contains considerable information. There are 
basically three qualitative behaviors exhibited for different $R_V$, 
indicated by the (colored) background shading. For $R_V \lesssim 100$, 
the laminar Roberts flow is hydrodynamically steady-state and laminar, 
but dynamo action is still possible for large enough $R_M$. For 
$100 \lesssim R_V \lesssim 1000$, Roberts flow treated purely 
hydrodynamically is temporally periodic but not turbulent. For 
$R_V \gtrsim 1000$, the Roberts flow develops a full turbulent 
spectrum hydrodynamically. In all three regimes, dynamo action is 
exhibited, but is different in the three regimes. The laminar 
regime was extensively studied in Ref. \cite{Feudel03}. Our definitions 
for the Reynolds numbers are different, but the results displayed in 
Figs. \ref{fig:crmm} and \ref{fig:crll} are consistent with 
previous results in the range $P_M=[0.5,1]$ if our Reynolds numbers 
are divided by $2\pi$ (corresponding approximately to the integral 
scale of the laminar flow).

The threshold curve connecting diamonds ($\Diamond$) is for the Roberts 
flow with $f=g$ (helical, but not maximally so). The segment connecting 
squares ($\Box$) is for $f=g/\sqrt{2}$ (maximal helicity). The segment 
connected by triangles ($\triangle$) is for $f=g/0.77$, a less helical 
flow than $f=g$. The threshold curve connecting crosses ($\times$) is 
the threshold curve for the Taylor-Green (TG) flow from Ref. 
\cite{Ponty05}. All of these are direct numerical simulation (DNS) 
results. (We regard the fact that the Taylor-Green curve and the Roberts 
flow curve with $f=g$ have a common region above $R_V \sim 100$ to be 
coincidental). The curve connecting crosses ($\times$) with a dashed 
line is the result from Ref. \cite{Ponty05} for the ``$\alpha$-model'', 
or Lagrangian averaged model, of MHD.

Noteworthy in Fig. \ref{fig:crmm} is the qualitative similarity of the 
behavior of the threshold curve between the Roberts flow and the TG 
results from Ref. \cite{Ponty05}: a sharp rise in $R_M^c$ with the 
increase in the degree of turbulence in the velocity field, followed 
by a plateau in which further increases in $R_V$ show little effect 
upon $R_M^c$. It must be kept in mind that for both situations, the 
amplitude of the forcing field ${\bf f}$ is being adjusted so that 
$U$ and the total kinetic energy remain approximately constant, 
even though $R_V$ is increasing. Whether such a procedure corresponds 
to a physical driving force must be decided on a case-by-case basis.

Figure \ref{fig:crll} shows the threshold curve for the Roberts flow 
with $f=g$ as a function of the inverse of the magnetic Prandtl 
number, $P_M^{-1}$. This curve shares some similarities with the TG flow, 
but also important differences. As in Ref. \cite{Ponty05}, between the 
laminar and turbulent regimes a sharp increase in $R_M^c$ is observed. 
Also, in the turbulent flow $R_M^c$ seems to be independent of the value 
of the magnetic Prandtl number. But while the TG force is not a solution 
of the Euler equations and was designed to generate smaller and smaller 
scale fluctuations as the Reynolds number $R_V$ is increased, the Roberts 
flow goes through several instabilities as $R_V$ is varied. As a result, 
the threshold for dynamo action in the $R_M$ vs. $P_M^{-1}$ plane is 
double-valued. For a given value of $P_M^{-1}$ two values of $R_M^c$ 
exist according to the hydrodynamic state of the hydrodynamic system, 
(e.g. laminar, periodic, or turbulent flow).

\begin{figure}
\includegraphics[width=9cm]{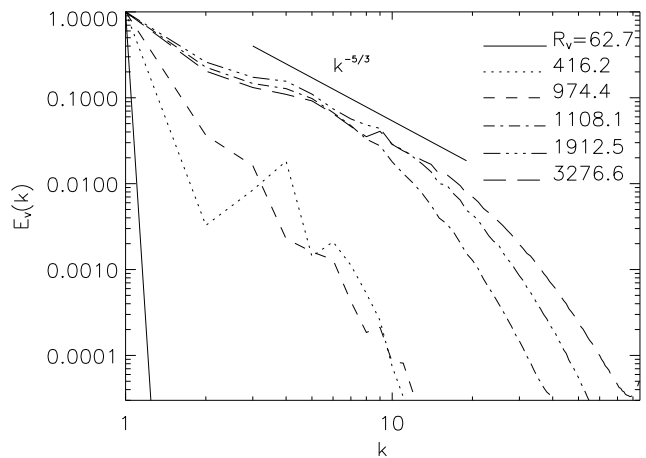}
\caption{Kinetic energy spectra as a function of $R_V$. The Kolmogorov's 
    spectrum is showed as a reference.}
\label{fig:kspec}
\end{figure}

\begin{figure}
\includegraphics[width=9cm]{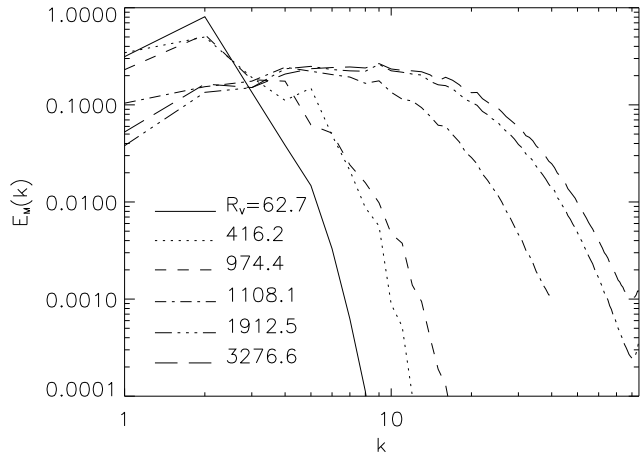}
\caption{Magnetic energy spectra during the kinematic regime, 
    for different values of $R_V$. The values of $R_M$ for each curve 
    correspond to the smallest value for which dynamo action was 
    observed (see Fig. \ref{fig:growth}).}
\label{fig:mspec}
\end{figure}

Figure \ref{fig:kspec} is a plot of the kinetic energy spectra for the 
values of $R_V$ shown in Fig. \ref{fig:crmm}, for $f=g$, normalized so 
that $E_V(k=1)$ is unity for all cases. This is done to display the 
gradual widening of the spectrum as $R_V$ increases. Figure \ref{fig:mspec} 
shows corresponding magnetic spectra, normalized somewhat differently: 
the energy contained in the interval $1 \le k \le 5$ is the same in all 
cases. This is done to emphasize the fact that the peak in the 
magnetic energy spectrum migrates to higher values as $R_V$ increases: 
the excited magnetic field develops more and more small-scale features. 
This may be related to the fact that because the forcing occurs at such 
low wavenumbers, inverse magnetic helicity cascades are effectively 
ruled out.

\begin{figure}
\includegraphics[width=9cm]{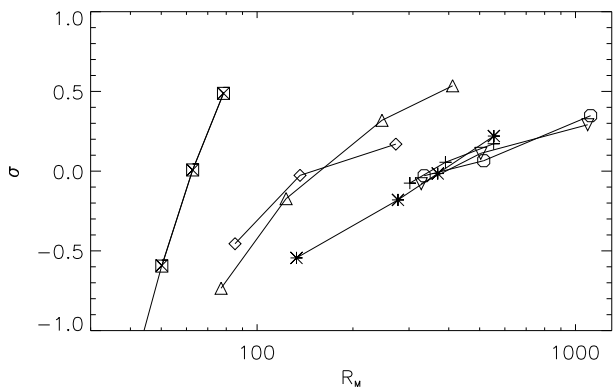}
\caption{Growth rates as a function of $R_M$. Each line corresponds 
    to several simulations at constant $R_V$ (fixed $\nu$), and each 
    point in the line indicates the exponential growth (or decay) rate 
    at a fixed value of $R_M$. The point where each curve crosses 
    $\sigma=0$ gives the threshold $R_M^c$ for dynamo instability. 
    $R_V = 62.7$ ($\Box$), $R_V = 125.5$ ($\times$), $R_V = 416.2$ 
    ($\triangle$), $R_V = 974.4$ ($\Diamond$), $R_V = 1108.1$ ($*$), 
    $R_V = 1327.7$ ($+$), $R_V = 1912.5$ ($\circ$), and $R_V = 3276.6$
    ($\bigtriangledown$).}
\label{fig:growth}
\end{figure}

Figure \ref{fig:growth} shows how the thresholds ($\sigma=0$) for the 
$R_M^c$ curves were calculated. For small initial $E_M$, broadly 
distributed over $k$, $\eta$ was gradually decreased in steps to 
raise $R_M$ in the same kinetic setting until a value of 
$\sigma \approx 0$ was identified. That provides a single point on 
such curves as those in Fig. \ref{fig:crmm}.

Each simulation at a fixed value of $\nu$ and $\eta$ (or fixed 
$R_V$ and $R_M$) was extended for at least 100 large-scale turnover 
times to rule out turbulent fluctuations and obtain a good fit to the 
exponential growth. All the simulations were well-resolved and 
satisfied the condition $k_\nu/k_{max}<1$, where 
$k_\nu=(\epsilon/\nu^3)^{1/4}$ is the Kolmogorov lengthscale, 
$\epsilon$ is the energy injection rate, $k_{max}=N/3$ is the largest 
resolved wavenumber, and $N$ is the linear resolution of the simulation. 
When this condition was not satisfied, the resolution $N$ was increased, 
from $N=64$ until reaching the maximum spatial resolution in this work of 
$256$ grid points in each direction, and a maximum mechanical Reynolds 
of $R_V=3276.6$.

\begin{figure}
\includegraphics[width=9cm]{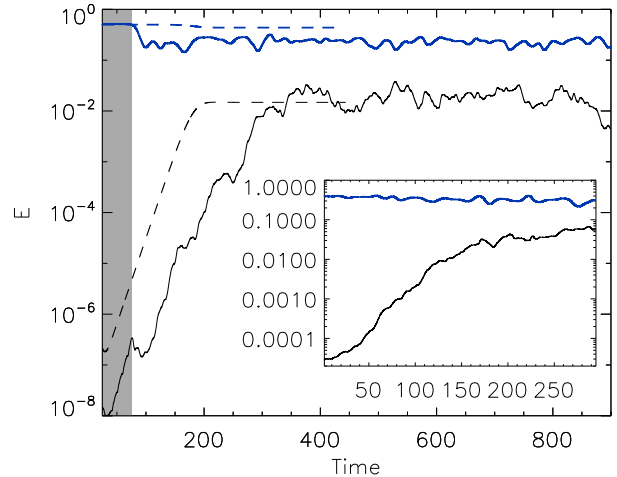}
\caption{(Color online) Time history of the total kinetic [thick 
    (blue) lines] and magnetic energy (thin lines) in dynamo 
    simulations. The dashed lines correspond to $R_V=62.7$ and 
    $R_M=78.4$ (laminar flow), while the solid lines are for 
    $R_V=R_M=416.2$. The shaded region indicates the period of 
    time when the flow is oscillating in this simulation. The 
    inset shows the time history for a turbulent run with 
    $R_V=3276.6$ and $R_M=1092.2$.}
\label{fig:ener}
\end{figure}

\begin{figure}
\includegraphics[width=9cm]{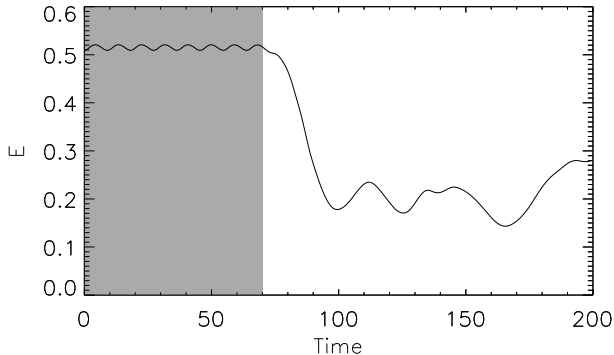}
\caption{Time history of the total energy in the dynamo simulation 
    with $R_V=R_M=416.2$. The shaded area is a blow up of the shaded 
    region in Fig. \ref{fig:ener} and corresponds to the hydrodynamic 
    oscillations.}
\label{fig:oscillation}
\end{figure}

Figure \ref{fig:ener} illustrates an interesting behavior that occurs 
when there is a transition from the laminar to the periodic regime 
of the Roberts flow ($f=g$). Figure \ref{fig:ener} shows the evolution 
of total kinetic energies $E_V$ and magnetic energies $E_M$ for 
$R_V = 62.7$ and $R_V = 416.2$. The flat part of the kinetic [thick 
(blue)] curve for $R_V = 416.2$ is characterized by small periodic 
oscillations too small to see on the logarithmic plot (they will 
be shown in Fig. \ref{fig:oscillation}). Meanwhile, the $E_M$ curve 
of magnetic energy is growing, somewhat irregularly. Rather suddenly, 
at about $t=70$, $E_V$ drops by more than a factor of 2 (see Fig. 
\ref{fig:oscillation}), and by $t \approx 300$ the magnetic energy 
has saturated at a level of about 1 per cent of the initial kinetic 
energy. Both fields oscillate irregularly after that, and are weakly 
turbulent. It is unclear how such a small magnetic excitation succeeds 
at shutting down such a large fraction of the flow. As will be 
shown later, this large drop is associated with the instability of the 
large scale flow. The inset shows the full time history of $E_V$ and 
$E_M$, for $R_V=3276.6$ and $R_M=1092.2$ when the turbulence is fully 
developed. The dashed line illustrates, for comparison, how simply the 
magnetic energy exponentiates and saturates in the laminar steady-state 
regime ($R_V=62.7$). Figure \ref{fig:oscillation} shows in detail the 
suppression of the flow, manifested as a drop in the {\it total} energy, 
at $t \approx 70$.

These oscillations between the hydrodynamic laminar and turbulent 
regime in the Roberts flow have been previously found by Feudel et al. 
\cite{Feudel03}. The authors pointed out that in this regime, close to 
the threshold $R_M^c$ the dynamo exhibits an intermittent behavior, 
with bursts of activity. The oscillatory flow is stable to small 
perturbations (e.g. numerical noise in the code), but as the magnetic 
energy grows the flow is perturbed by the Lorentz force and goes to a 
weakly turbulent regime. As noted in Ref. \cite{Feudel03}, if 
$R_M$ is close to $R_M^c$ then the magnetic field decays, the flow 
relaminarizes and the process is repeated. However, as observed in Fig. 
\ref{fig:ener}, if $R_M$ is large enough the weakly turbulent flow can 
still excite a dynamo, and the magnetic field keeps growing exponentially 
until reaching the non-linear saturation even after the hydrodynamic 
instability takes place.

\begin{figure}
\includegraphics[width=9cm]{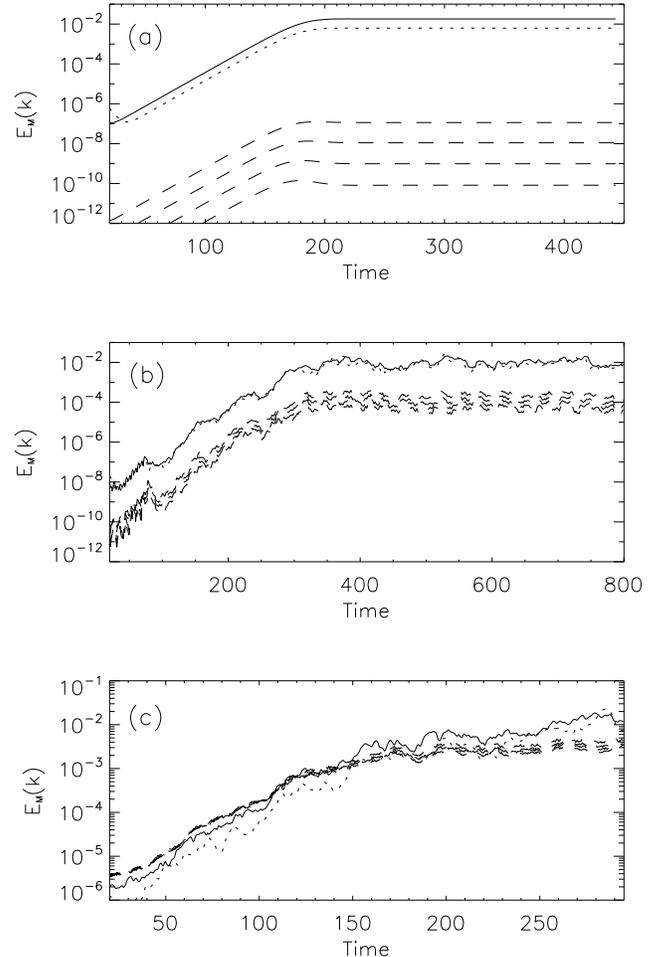}
\caption{Evolution of the magnetic energy in different shells in Fourier 
    space: (a) $R_V=62.7$ and $R_M=78.4$ (laminar flow), (b) 
    $R_V=R_M=416.2$ (periodic case), (c) $R_V=3276.6$ and $R_M=1092.2$ 
    (turbulent regime). The dotted line corresponds to $k=1$, solid 
    line to $k=2$, and the dashed lines to $k=9,10,11,12$.}
\label{fig:modes}
\end{figure}

Figure \ref{fig:modes}(a) shows the temporal growth of several Fourier 
components of the magnetic field in the laminar regime ($R_V=62.7$). 
A straightforward exponentiation, followed by a flat, steady-state, 
leveling-off exhibits the same growth rate for all harmonics. This 
indicates the existence of a simple unstable normal mode which 
saturates abruptly near $t \approx 180$. The behavior is much noisier 
for $R_V=416.2$ and $3276.6$ as shown in Figs. \ref{fig:modes}(b) and 
\ref{fig:modes}(c). Note that in the simulation with $R_V=416.2$, 
for $t<70$ all the magnetic modes oscillate with the same frequency 
as the hydrodynamic oscillations. In Figure \ref{fig:modes}, the 
dotted line and solid line above are, respectively, for $k=1$ and 
$k=2$. The remaining four are for $k=9$ through 11. The modes in 
between occupy the open space in between more or less in order. The 
same modes are shown for $R_V=416.2$ in Fig. \ref{fig:modes}(c), which 
illustrates a broad sharing of $E_B$ among many modes and a consequent 
excitation of small-scale magnetic components.

\begin{figure*}
\includegraphics[width=12cm]{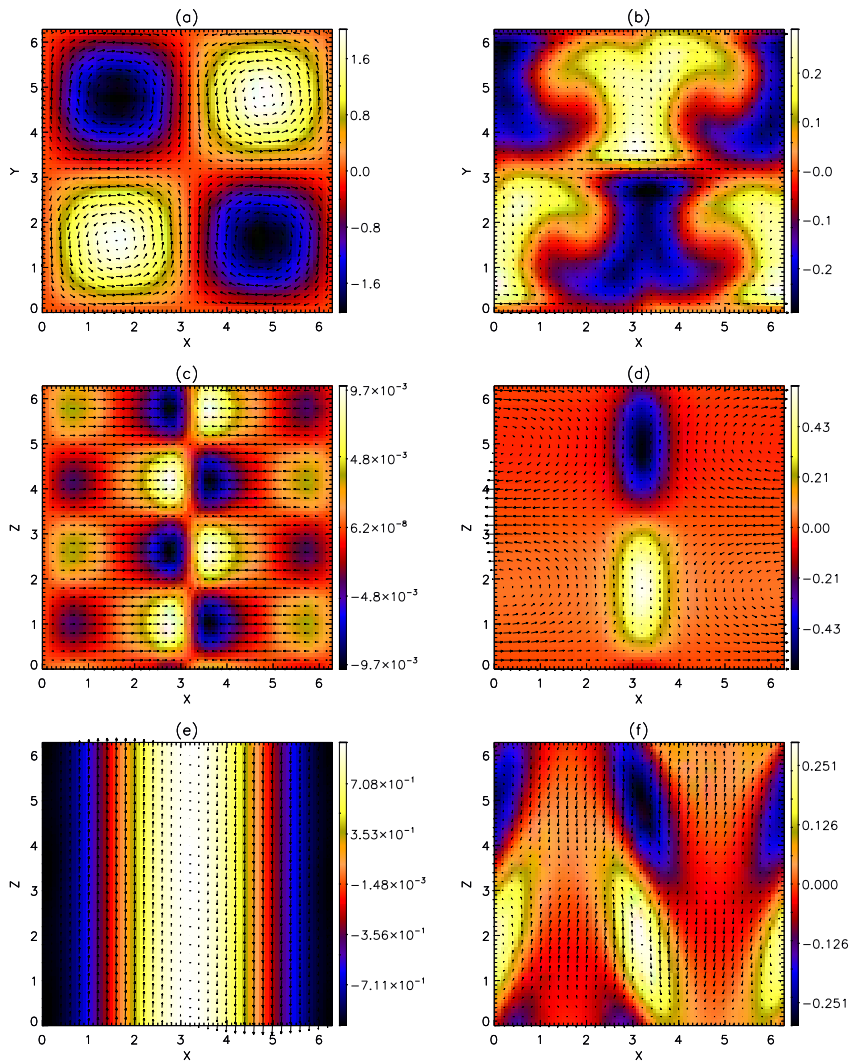}
\caption{(Color online) Plots of the kinetic and magnetic fields for 
    the saturated regime of the run with $R_V=62.7$ and $R_M=78.4$: 
    (a) cut at $z=0$, $vz$ in color and $vx$, $vy$ indicated by 
    arrows, (b) same as in (a) for the magnetic field, (c) cut at 
    $y=0$, $vy$ in color and $vx$, $vz$ indicated by arrows, (d) same 
    as in (c) for the magnetic field, (e) same as in (b) but for a cut 
    at $y=\pi/4$, and (f) same as in (e) for the magnetic field.}
\label{fig:color1}
\end{figure*}

\begin{figure*}
\includegraphics[width=12cm]{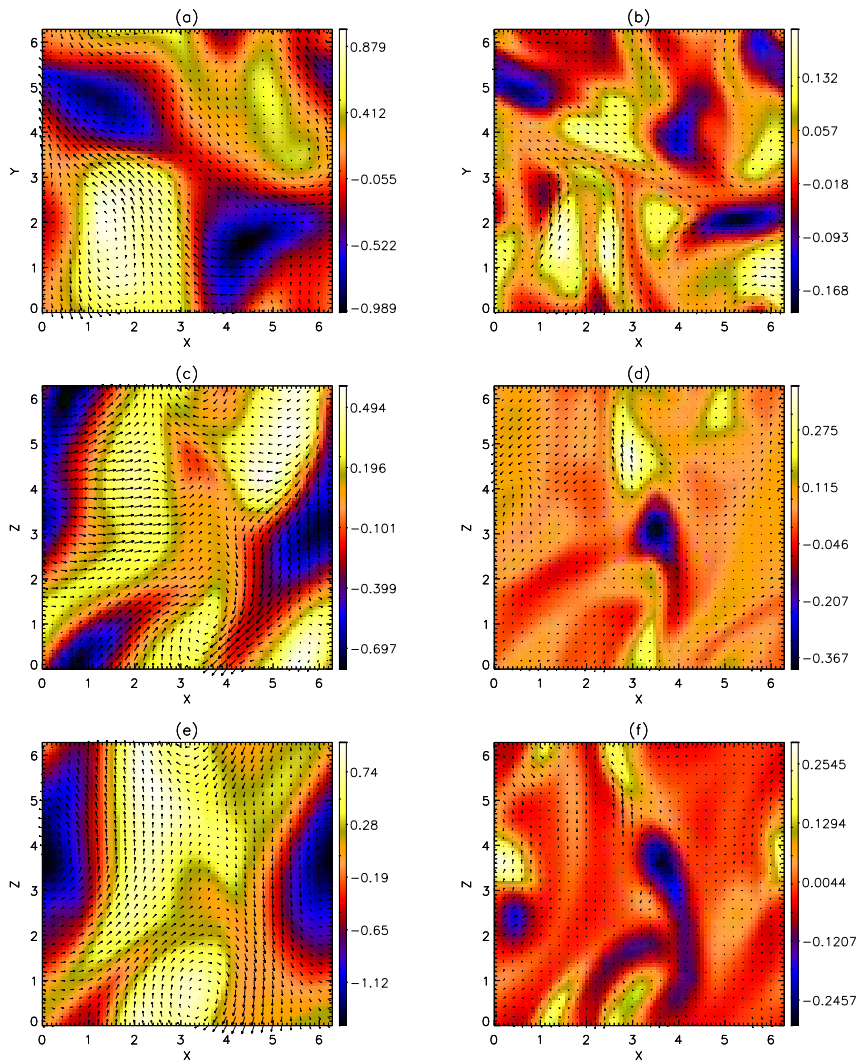}
\caption{(Color online) Plots of the kinetic and magnetic fields for 
    the saturated regime of the run with $R_V=R_M=416.2$. Labels and 
    fields are as in Fig. \ref{fig:color1}.} 
\label{fig:color2}
\end{figure*}

\begin{figure*}
\includegraphics[width=12cm]{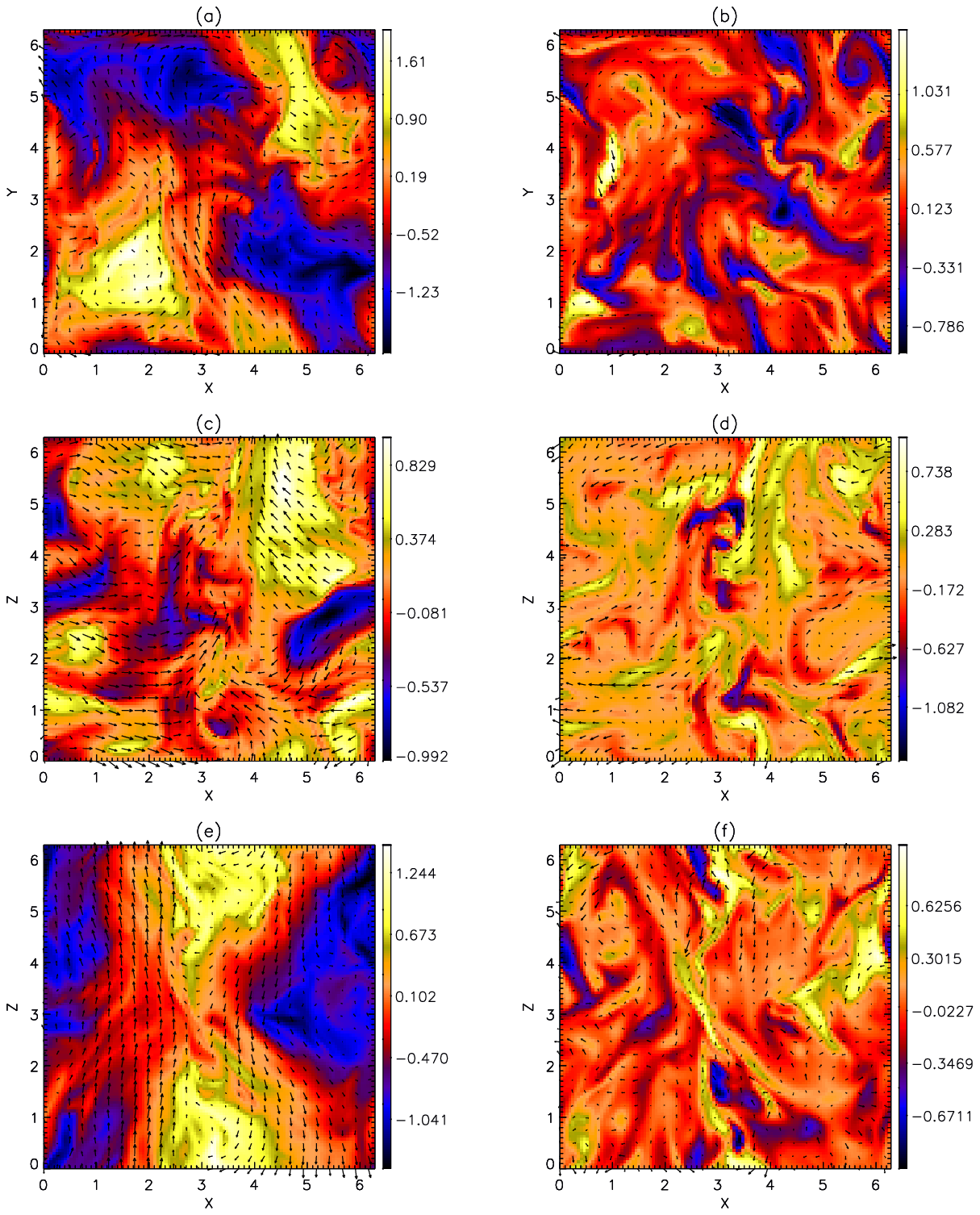}
\caption{(Color online) Plots of the kinetic and magnetic fields for 
    the saturated regime of the run with $R_V=3276.6$ and $R_M=1092.2$. 
    Labels and fields are as in Fig. \ref{fig:color1}.} 
\label{fig:color3}
\end{figure*}

Plots of the kinetic and magnetic fields are shown in Figs. 
\ref{fig:color1}. The left column shows the velocity field in the 
saturated state for $R_V=62.7$, and the right column shows the 
magnetic field at the same time. The arrows indicate the vector 
components in the planes shown and the colors indicate the strengths 
of the perpendicular components. Figures \ref{fig:color1}(a) and (b) 
are for the plane $z=0$ and Figs. \ref{fig:color1}(c) and (d) are for 
the plane $y=0$. Figs. \ref{fig:color1}(e) and (f) are for the plane 
$y=\pi/2$. The velocity configuration shown in Fig. \ref{fig:color1}(a) 
is quite similar to the way it looks at $t=0$, but the $z$-dependences 
apparent in Figs. \ref{fig:color1}(c), (d), and (f) are not present in 
the initial flow.

Figs. \ref{fig:color2} are similar color plots for the saturated 
regime for $R_V=416.2$. All the same quantities are being 
displayed at the same planes as in Figs. \ref{fig:color1}. The 
initial conditions are no longer recognizable in the saturated 
state, but is not yet sufficiently disordered that one would be 
forced to call it ``turbulent''. Moreover, note that the 
four ``cells'' characteristic of the laminar Roberts flow 
[Fig. \ref{fig:color1}(a)] are not present in this late stage 
of the dynamo. During the early kinematic regime, when the 
hydrodynamic oscillations are observed, a slightly deformed 
version of these cells can be easily identified in the flow (not 
shown). When the magnetic energy grows due to dynamo action, the 
flow is unable to maintain this flow against the perturbation of 
the Lorentz force. This causes the large-scale flow to destabilize, 
and the kinetic energy in the shell $k=1$ drops by a factor of two. 
This instability of the large-scale modes is associated with the 
large drop of the kinetic and the total energy at $t \approx 70$ 
(Fig. \ref{fig:oscillation}).

\begin{figure}
\includegraphics[width=9cm]{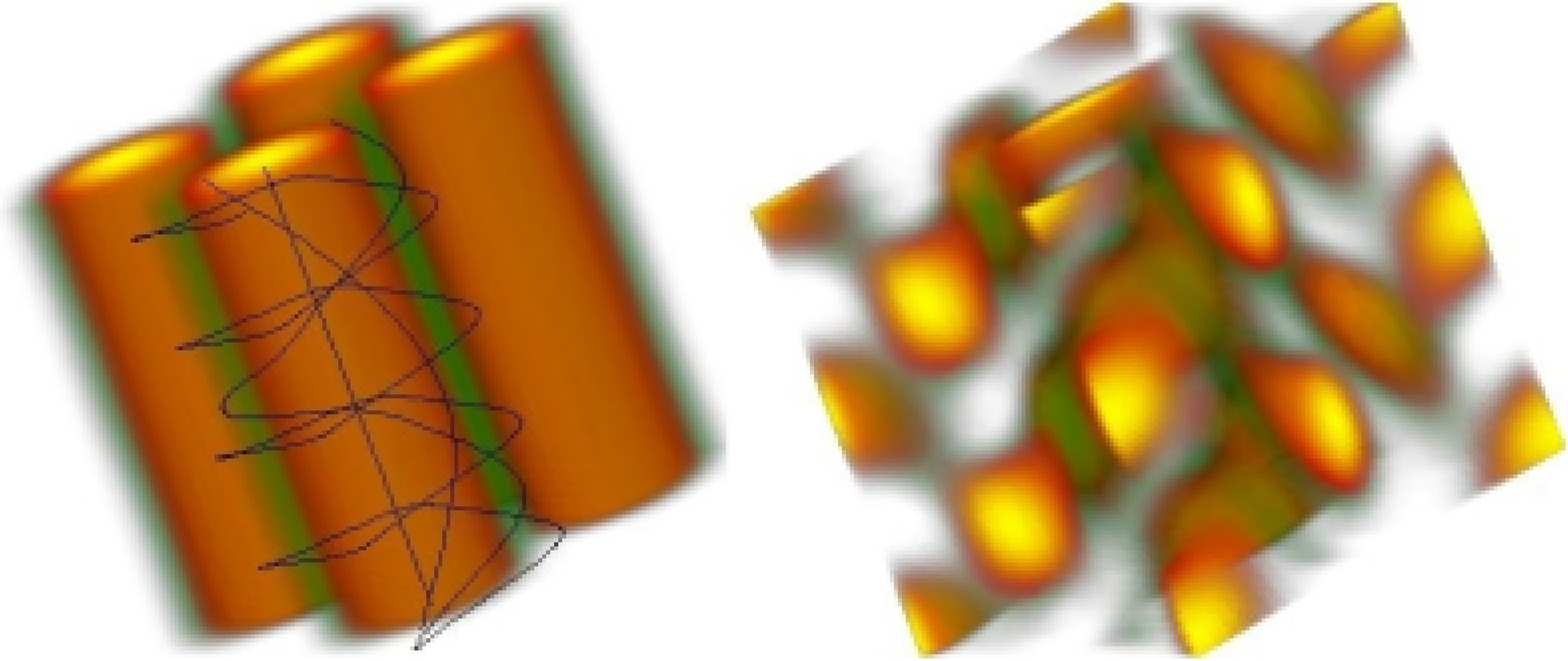}
\caption{(Color online) Visualization of the kinetic (left) and 
    magnetic energy density (right) for the saturated regime of 
    the run with $R_V=62.7$ and $R_M=78.4$. Velocity field 
    lines are indicated in black.} 
\label{fig:render1}
\end{figure}

\begin{figure}
\includegraphics[width=9cm]{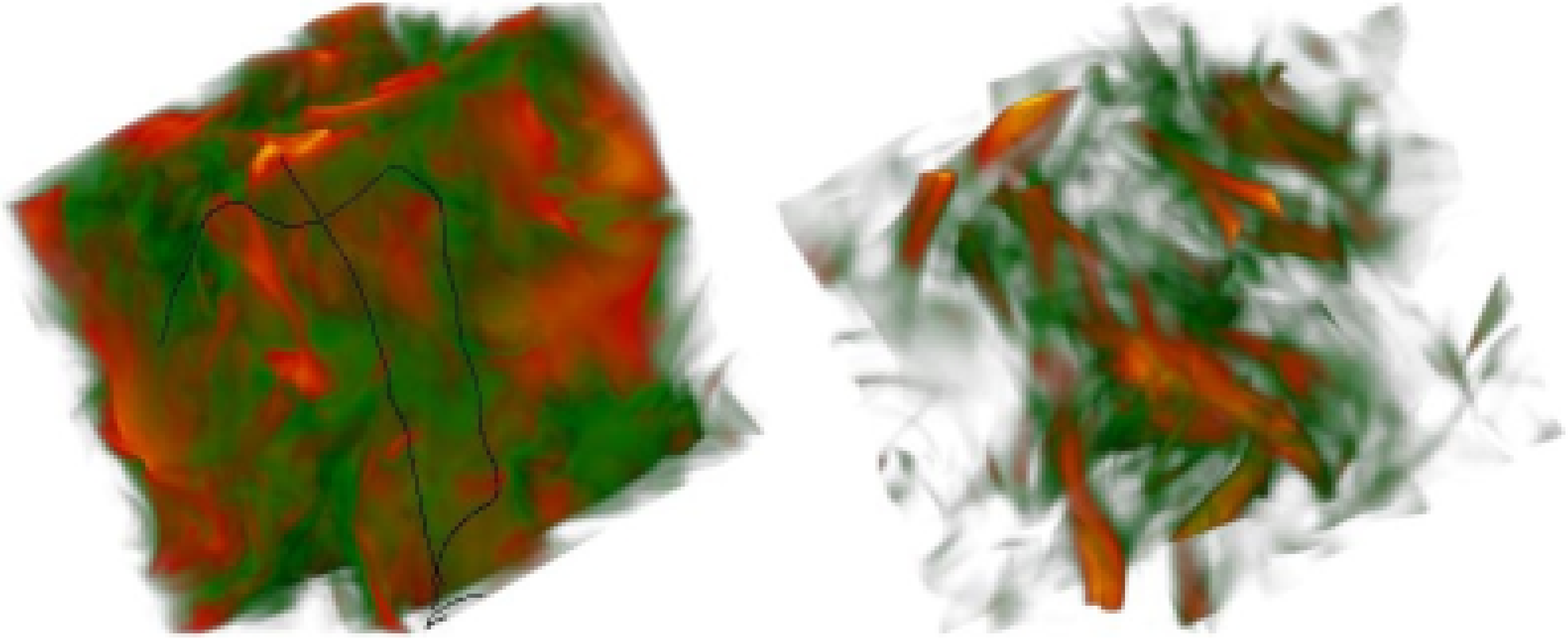}
\caption{(Color online) Visualization of the kinetic (left) and 
    magnetic energy density (right) for the saturated regime of 
    the run with $R_V=3276.6$ and $R_M=1092.2$. Velocity field 
    lines are indicated in black.} 
\label{fig:render2}
\end{figure}

By contrast, the same fields are exhibited in the same planes 
in Figs. \ref{fig:color3} in the saturated regime for $R_V=3276.6$. 
Here the truly turbulent nature of the flow is now apparent, 
particularly in the highly disordered magnetic field plots in the 
right-hand column.

\begin{figure}
\includegraphics[width=9cm]{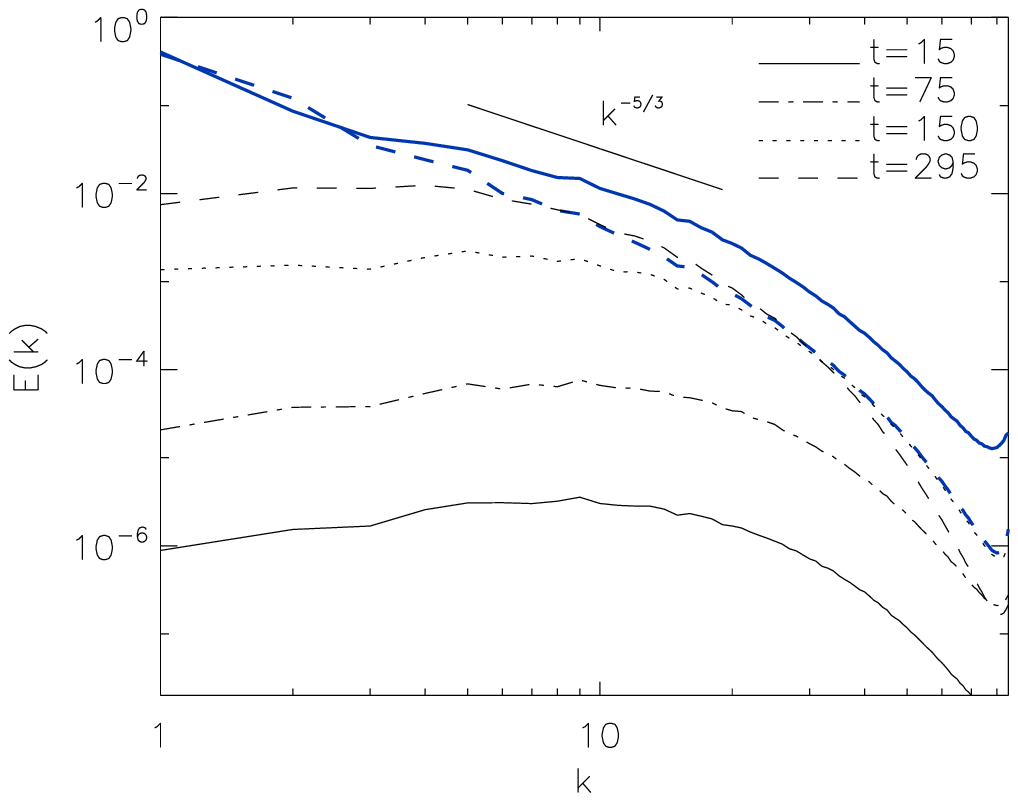}
\caption{(Color online) Kinetic [thick (blue) line] and magnetic 
    energy spectra for different times for the simulation with 
    $R_V=3276.6$ and $R_M=1092.2$.} 
\label{fig:spectrum}
\end{figure}

Figure \ref{fig:render1} is a three-dimensional perspective plot 
of the kinetic and magnetic energy density for $R_V=62.7$ at a 
late time in the saturated regime. The kinetic energy distribution 
(on the left) is not much different than it was at $t=0$. The 
helical properties of the Roberts flow can be directly appreciated 
in the field lines indicated in black. In this regime, the flow is 
still laminar as previously indicated. The magnetic field is 
stretched and magnetic energy amplified in the four helical tubes, 
and then expelled out of the vortex tubes, accumulating in the 
stagnation points \cite{Roberts72,Feudel03}. Since the velocity 
field has no dependence in the $z$-direction, the magnetic 
field that can be sustained by dynamo action has to break this 
symmetry and displays a clear periodicity in this direction. The 
same energy densities are exhibited at a late time for the case of 
$R_V=3276.6$ in Fig. \ref{fig:render2}, and the highly filamented 
and disordered distributions characteristic of the turbulent 
regime are again apparent. Note however that still some 
helicity can be identified in the velocity field lines shown.

In Ref. \cite{Mininni05a} a suppression of small scale turbulent 
fluctuations and an evolution of the system to a state with effective 
magnetic Prantdl number of order one was observed in the nonlinear 
saturation of the turbulent dynamo. Here a similar effect is observed, 
although the suppression of small scales is weaker probably due to the 
presence of the external forcing at $k \approx 1$ which does not leave 
room for a large scale magnetic field to develop. Figure \ref{fig:spectrum} 
shows the time evolution of the kinetic and magnetic energy spectra in 
the run with $R_V=3276.6$ and $R_M=1092.2$. While at early times the 
magnetic energy spectrum peaks at small scales ($k \approx 9$), at late 
times the magnetic spectrum is flat for small $k$ and drops together with 
the kinetic energy. The kinetic spectrum is strongly quenched and has 
a large drop at small scales.

\section{\label{sec:summary}SUMMARY AND DISCUSSION}

One apparent outcome of these computations has been to confirm 
the intuitive impression that dynamo amplification of very small 
magnetic fields in conducting fluids is easier if mechanical helicity 
is present. This is true in velocity fields which are both turbulent 
and laminar. The values of $R_M^c$ which are the lowest found 
($\sim 10$) are well below those in several existing experimental 
searches.

It is also somewhat reassuring to find that the qualitative behavior 
of dynamo thresholds with decreasing viscosity (increasing Reynolds 
number at fixed $U$) is as similar as it is to that found for the 
non-helical TG flow in Ref. \cite{Ponty05}. In particular, since 
the simulations discussed here were forced at almost the largest scale 
available in the periodic domain, a turbulent regime for $P_M<1$ where 
$R_M^c$ is approximately independent of $P_M$ was reached using only DNS, 
while for the TG flow two different models \cite{Mininni05b,Ponty04} 
for the small scales were needed. The similarities in the behavior of 
the threshold for the two flows for $P_M$ small enough brings more 
confidence to the ability of subgrid scale models of MHD turbulence 
to predict results in regimes of interest for astrophysics and 
geophysics that are today out of reach using DNS. 
That being said, it should be admitted that the Roberts flow in a way 
exhibits a richer set of possibilities in that the dynamo activity is 
somewhat different in each of the three regimes (laminar and steady-state, 
oscillatory, and turbulent).

Dynamo action is to be regarded as of many types \cite{Mininni05a} and 
situation-dependent. The forms of the magnetic fields developed and 
their characteristic dimensions are determined to a considerable extent 
by the mechanical activity that excites them and by the geometric 
setting in which they take place. If it is desired to apply the 
theoretical and computational results to planetary dynamos or 
laboratory experiments, then rectangular periodic conditions appear 
to be a constraint that should be dispensed with as soon as feasible.

\begin{acknowledgments}
The authors are grateful for valuable comments to Dr. Annick Pouquet. 
The NSF grants ATM-0327533 at Dartmouth College and CMG-0327888 at 
NCAR supported this work in part and are gratefully acknowledged. 
Computer time was provided by the National Center for Atmospheric 
Research.
\end{acknowledgments}

\bibliography{ms}

\end{document}